\def\bea{\begin{eqnarray}}
\def\eea{\end{eqnarray}}

\documentclass[aps,pre,twocolumn,superscriptaddress,amsmath,amssymb,longbibliography]{revtex4-1}

\usepackage{graphicx}

\usepackage{dcolumn}

\usepackage{enumitem,kantlipsum}

\usepackage{bm}
\usepackage[normalem]{ulem}
\usepackage{color}
\usepackage{gensymb}

\usepackage{epstopdf}
\newcommand{\bew}{\begin{widetext}}
\newcommand{\ew}{\end{widetext}}

\newcommand{\bz}{\mathbf{z}}

\newcommand{\bq}{\mathbf{q}}

\newcommand{\bv}{\mathbf{v}}

\newcommand{\br}{\mathbf{r}}
\newcommand{\brp}{\mathbf{r}_{_\parallel}}
\newcommand{\bR}{\mathbf{R}}

\newcommand{\hx}{\hat{{\bf x}}}

\newcommand{\hp}{\hat{{\bf p}}}

\newcommand{\sep}{ \ \ \ , \ \ \ }

\newcommand{\beq}{\begin{equation}}
\newcommand{\eeq}{\end{equation}}
\newcommand{\beqn}{\begin{eqnarray}}
\newcommand{\eeqn}{\end{eqnarray}}

\newcommand{\stkout}[1]{\ifmmode\text{\sout{\ensuremath{#1}}}\else\sout{#1}\fi}

\definecolor{green}{rgb}{0,0.5,0}

\newcommand{\cgreen}{\color{green}}

\begin{document}
\title{Swarming bottom feeders: Flocking at solid-liquid interfaces}
\author{Niladri Sarkar}\email{niladri2002in@gmail.com}
\affiliation{Instituut-Lorentz, Leiden University, P.O. Box 9506, 2300 RA Leiden, The Netherlands}
\author{Abhik Basu}\email{abhik.123@gmail.com, abhik.basu@saha.ac.in}
\affiliation{Condensed Matter Physics Division, Saha Institute of
Nuclear Physics, Calcutta 700064, West Bengal, India} 
\author{John Toner}\email{jjt@uoregon.edu}
\affiliation{Department of Physics and Institute of Theoretical Science, University of Oregon, Eugene, Oregon 97403, USA}

\date{\today}
\begin{abstract}
 We present the hydrodynamic theory 
  of coherent collective motion (``flocking'') at a solid-liquid interface, and many of its predictions for experiment.  
We find that such systems are stable, and have long-range orientational order, over a wide range of parameters. When stable, these systems exhibit ``giant number fluctuations'', which  grow
as the 3/4th power of the
mean  number. Stable systems also exhibit anomalous rapid diffusion of tagged particles suspended in the passive fluid along any directions in a plane parallel to the solid-liquid interface,  whereas the diffusivity along the direction perpendicular to  the plane is not anomalous. In  the remaining parameter space, the system becomes unstable. 
\end{abstract}

\maketitle
 

Many ``active'' systems consist of macroscopically large numbers of self-propelled particles that align their directions of motion. This occurs both in  
living~\cite{kruse04,kruse05,goldstein13,saintillan08,hatwalne04} and 
synthetic~\cite{saha14,cates15,narayan07,lubensky09,marchetti08}  systems. Such ``active orientationally ordered phases'' exhibit many phenomena impossible in their equilibrium analogs (e.g., nematics~\cite{deGennes}), including spontaneous breaking of continuous symmetries in two dimensions\cite{vicsek95, tonertu95, toner98, toner05}, instability in the extreme Stokesian limit\cite{simha2002}, 
and giant number fluctuations~\cite{Chate+Giann, toner2019giant, ramaswamy03}.


``Dry'' active systems  - i.e., those lacking momentum conservation due to, e.g.,  friction with a substrate~\cite{wolgemuth2002, toner98,ramaswamy03}- behave quite differently from ``wet'' active fluids (i.e., those {\it with} momentum conservation)~\cite{lushi2014}.  

\begin{figure}[htb]
\includegraphics[height=4.1cm,width=7cm]{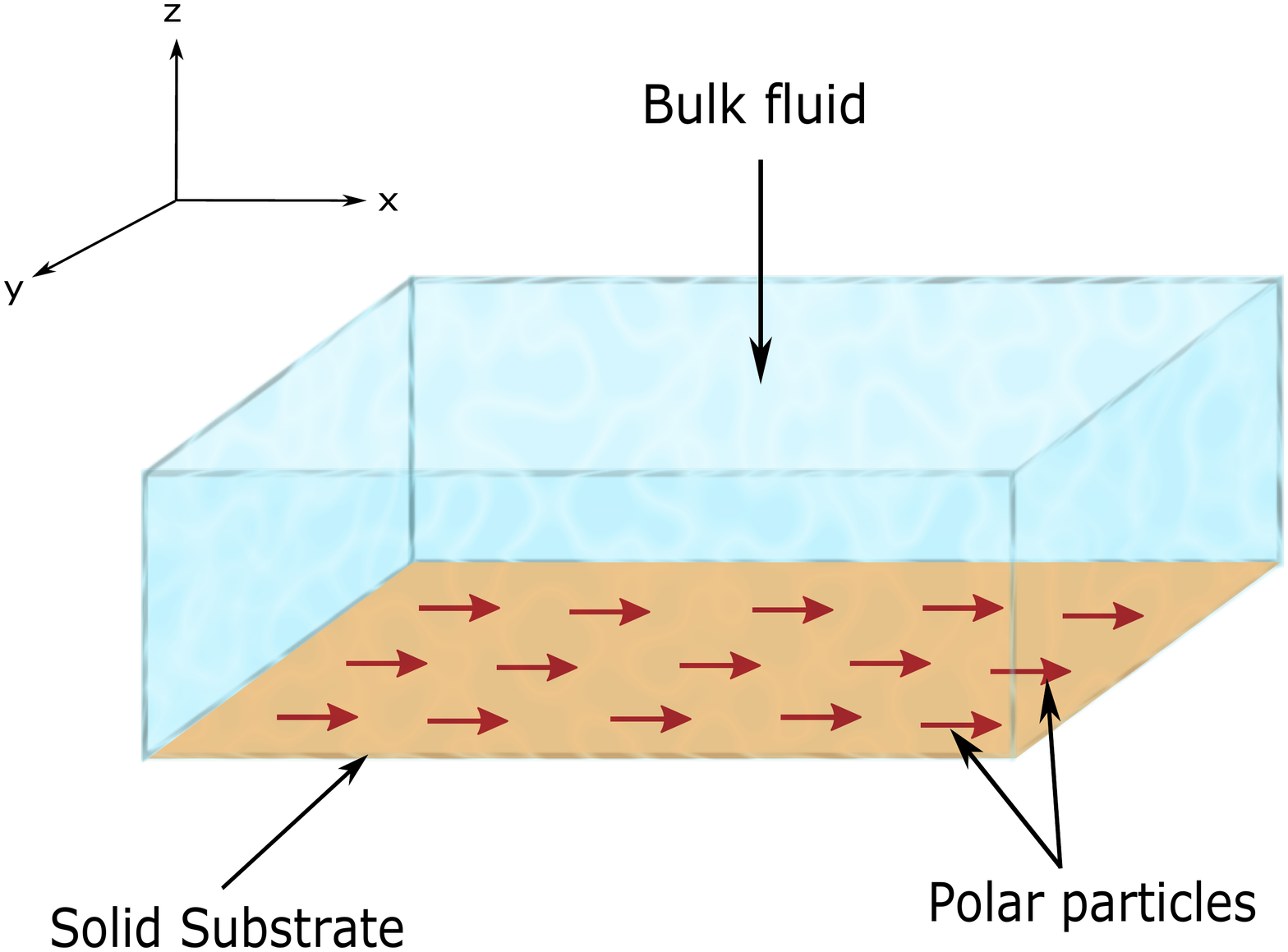}
\caption{(Color online) Schematic diagram of our system: a layer of active polar particles 
moving  on a solid substrate with a passive ambient (``bulk'') fluid above.}
\label{schem}
\end{figure}

In this paper, we  present 
the first 
theory of 
a natural hybrid of these two cases: 
polar active particles at a solid-liquid interface (see figure (\ref{schem})). We are motivated by experiments 
~\cite{schaller13} in which highly concentrated actin filaments on a solid-fluid interface are propelled by  motor proteins, and those   of Bricard {\em et al} ~\cite{bricard2013, Geyer17}, who studied the emergence of macroscopically directed motion in ``Quincke rollers''. The latter are 
motile colloids, 
spontaneously rolling on a solid substrate 
when a sufficiently strong electric field is applied.


These systems differ  from both dry and  wet active matter, as defined above, by having both 
friction from the underlying solid substrate and the long range 
hydrodynamic interactions due to the overlying bulk passive fluid. 


The geometry we consider here, as in Ref.~\cite{schaller13, bricard2013}, places a collection of polar, self-propelled particles 
 at the flat interface (
 the $x$-$y$ plane of our coordinate system) between a solid 
substrate and a   semi-infinite bulk isotropic  and incompressible passive liquid. 
as illustrated in Fig.~\ref{schem}. 
 We   consider
the  extreme Stokesian limit,  in which 
inertial forces are
completely negligible compared to viscous forces.

  The most surprising result    of our work is that,  {\it even in the presence of noise,} this system can   be in a stable, long-range ordered polar state, in sharp contrast to ``wet'' active systems, which are generically unstable 
\cite{simha2002} at low Reynolds number, and equilibrium systems, which cannot display long range orientational order in two dimensions   at finite temperature\cite{MW, xtalfoot, 2dxtal,teth}.  

Remarkably,  this ordered state is predicted even by a  linear  theory. Furthermore, this linear theory provides
an asymptotically exact long wavelength description, in contrast to dry polar active systems, which can only be correctly described by a non-linear theory. Indeed, dry polar active systems  can only exhibit long range order due to non-linear effects \cite{vicsek95, tonertu95, toner98, toner05}.

 Concomitant with the long-range polar order, the density fluctuations are {\em giant}:  the 
standard deviation $\sqrt{\langle(N-\langle N\rangle)^2\rangle}$ of the number $N$ of the active particles  contained in a fixed open area scales with its average $\langle N\rangle$ according to
\beq
\sqrt{\langle(N-\langle N\rangle)^2\rangle}\propto\langle N\rangle^{3/4} \,.
\label{gnf}
\eeq
This agrees very well with the  experiments of ~\cite{schaller13},  which found $\sqrt{\langle(N-\langle N\rangle)^2\rangle}\propto\langle N\rangle^{0.8}$. Note that our prediction should not be confused with qualitatively similar predictions for dry active matter \cite{Chate+Giann,GNF} and active nematics \cite{AN}, for which the exponent is different, because they belong to  different universality classes. 

We also find that the fluctuations in the active fluid layer stir the bulk fluid above it,   making the diffusion  of a   passive tagged particle {\it parallel} to the active fluid layer {\em anomalous}:   specifically,  the mean squared displacement 
grows with time $t$ as $t\ln t$, whereas the diffusive motion 
{\it perpendicular } to the active fluid layer remains conventional, i.e., the mean squared displacement scales like $t$.  

 To understand the physics of this system, we   have constructed a theory which, when linearized for small fluctuations about a uniform reference state, is asymptotically exact in the long wavelength limit, and gives the above results.  We define $\hp(\brp,t)$ as the coarse grained polarization of the active particles, and $\rho(\brp,t)$ as the conserved  areal density  of the active polar particles on the surface. Taking our 
uniform reference state 
  to be $\hp(\brp,t) = \hx$ (see Fig.~\ref{schem}), and  $\rho=\rho_0$,    one hydrodynamic variable   is the transverse 
fluctuations $p_y$ of 
$\hp(\brp,t)$,   which we take to have unit magnitude, i.e., 
$|\hp|^2=1$. This is a non-conserved broken symmetry -  i.e., ``Goldstone" - mode. Our second hydrodynamic variable is  the 
fluctuations $\delta\rho(\brp,t)\equiv \rho(\brp,t)-\rho_0$   of the density from its mean value. 

These variables couple to the bulk passive fluid velocity $\bv(\brp, z, t)$ via an active boundary condition given below in \eqref{activebc}. Eliminating  $\bv(\brp, z, t)$ by solving the Stokes equation for the bulk fluid subject to this active boundary condition 
gives the  equations of motion for the spatially Fourier transformed fields $p_y(\bq,t)$ and $\delta \rho(\bq,t)$: 
\bea
\partial_t\delta\rho(\bq,t) =- iv_\rho [q_x\delta\rho(\bq,t)+\rho_c q_yp_y] + i\bq\cdot{\bf f}_\rho(\bq,t)\,,\nonumber\\
 \label{rholinprl}
\eea
\bew
\bea
\partial_tp_y(\bq,t) &=&- iv_pq_xp_y(\bq,t) -\gamma\left({q^2+q_y^2 \over q}\right)p_y (\bq,t)
-\left({\gamma_\rho\over\rho_c}\right)\left({q_xq_y \over q}\right)
\delta\rho(\bq,t)
-i\sigma_t q_y\delta\rho(\bq,t) + f_y(\bq,t) \,, 
\label{pyfin}
\eea
\ew
 where   $v_\rho$, $v_p$, $\gamma$, $\gamma_\rho$, $\rho_c$, and $\sigma_t$ are 
 parameters   of our model. 
 Note the non-analytic character of the   damping $\gamma$ and $\gamma_\rho$ terms in \eqref{pyfin};
   due to long-ranged hydrodynamic interactions 
mediated by the bulk passive fluid.

In \eqref{rholinprl} and \eqref{pyfin}, ${\bf f}_\rho$ and $f_y$ are zero-mean Gaussian white noises whose variances are parameters of our model.

 For stability, fluctuations must decay for all directions of $\bq$. We show in the  associated long paper (ALP)~\cite{alp} that this condition is satisfied provided that the analogs of the bulk compressibility and the shear and bulk viscosities in our system are all positive, and that  the coupling of the density of the active particles to their self-propelled speeds is not too strong. 

Thus, in contrast to ``wet'' active matter in the ``Stokesian'' limit \cite{simha2002, toner05}, our ``mixed'' system can be generically stable. Indeed, the requirements for stability are almost as easily met for these systems as for an equilibrium fluid. Furthermore, when the stability conditions are met,  fluctuations about the uniform ordered state in this model decay with a rate that scales linearly with $q$, quite different from the linear theory of dry active matter. The also propagate nondispersively with a wavespeed independent of $q$.



 This unusual damping in this linear theory is responsible for many novel phenomena: most strikingly,  it  makes $\langle p_y^2({\bf r}_\perp,t)\rangle$ asymptotically independent of the lateral size of the system, a tell-tale signature of orientational long-range order. It also leads
 to giant number fluctuations of the active particles given by (\ref{gnf}),   as mentioned earlier.
 

In the ordered state, the active particles ``stir" the passive fluid above them.
The  mean squared components $\langle v_x^2(\brp,z  ,t)\rangle$, and  $\langle v_y^2(\brp,z  ,t)\rangle$ of the   passive fluid velocity field $\bv(\brp, z ,t)$ thereby induced     are inversely proportional to the distance $z$ from the solid-fluid interface.  


  The unequal-time  correlations $\langle v_{x,y}(\brp,z,t)v_{x,y}(\brp,z,0)\rangle$ of the in-plane velocity fluctuations   of the passive fluid also exhibit long temporal correlations, which decay as $1/t$, 
whereas   the correlation 
$\langle v_z(\brp,z,t)v_z(\brp,z,0)\rangle$ of the 
the bulk fluid velocity   perpendicular to the surface  decays as $1/t^3$.

 The correlations  of the in-plane velocity in turn lead to anomalous diffusion of neutrally buoyant passive particles in the $x$- and 
 $y$-direction, with variances of the displacements  growing faster  with time than the linear dependence found  for simple brownian particles. Specifically, we find, for a particle that is initially a height $z_0$ above the solid-liquid interface,: 
\begin{eqnarray}
&&\langle(r_i(t)-r_i(0))^2\rangle =\left\{
\begin{array}{ll}
2D_it\left[\ln\left({v_0t\over z_0}\right)+O(1)\right]\,\,, \,\,\,\,\,\, t\ll{z_0^2\over D_z} \,,\\\\
D_it\left[\ln\left({v_0^2t\over D_z}\right)+O(1)\right]\,\,, \,\,\,\,\,\, t\gg{z_0^2\over D_z} \,,
\end{array}
\right.\nonumber\\
\label{anomdifshort}
\end{eqnarray}
where $i=x,y$, $v_0$ is a system-dependent characteristic speed 
(roughly speaking,  the  self-propulsion speed   of the active particles),  $z_0$ is the initial distance  from the surface,
and $D_{x,y,z}$ are diffusion constants which are independent of $z_0$.   Note that the mean square displacements depend on the initial height $z_0$ for  short times $t\ll z_0^2/D_z$,  but not   for long times $t\gg z_0^2/D_z$.
The 
  crossover between these limits is the time $t=z_0^2/D_z$ it takes for a neutrally buyoant particle to diffuse a distance $z_0$ in the $z$-direction.

Diffusion in the $z$-direction remains conventional, controlled by a $z$-independent diffusivity. 
 
 This set of predictions could also be tested experimentally by particle tracking of neutrally buoyant    tracer particles in the passive fluid.

Particles denser than the passive fluid, which therefore sediment, will also be affected by this activity induced flow. We find that particles sedimenting   at a speed $v_{\rm sed}\ll v_0$ from an initial height $z_0$ will, when they reach the surface, be spread out 
over a region of   RMS dimensions $\sqrt{\langle (x(z=0)-x(z=z_0))^2\rangle}$ and $\sqrt{\langle (y(z=0)-y(z=z_0))^2\rangle}$ in the $x$ and $y$ directions, respectively, with
\beq
\langle(r_i(t)-r_i(0))^2\rangle =2D_i\left({z_0\over v_{\rm sed}}\right)\ln\left({v_0\over v_{\rm sed}}\right) \sep v_{\rm sed}\ll v_0 \,, 
\label{anomsedx}
\eeq
where $v_0$ is roughly the mean speed of the active particles, and $v_{\rm sed}$ is  the speed at which the sedimenting particles sink.

Once again, these predictions should be readily testable in particle tracking experiments.

\begin{figure}[htb]
\includegraphics[height=5cm,width=3.7cm]{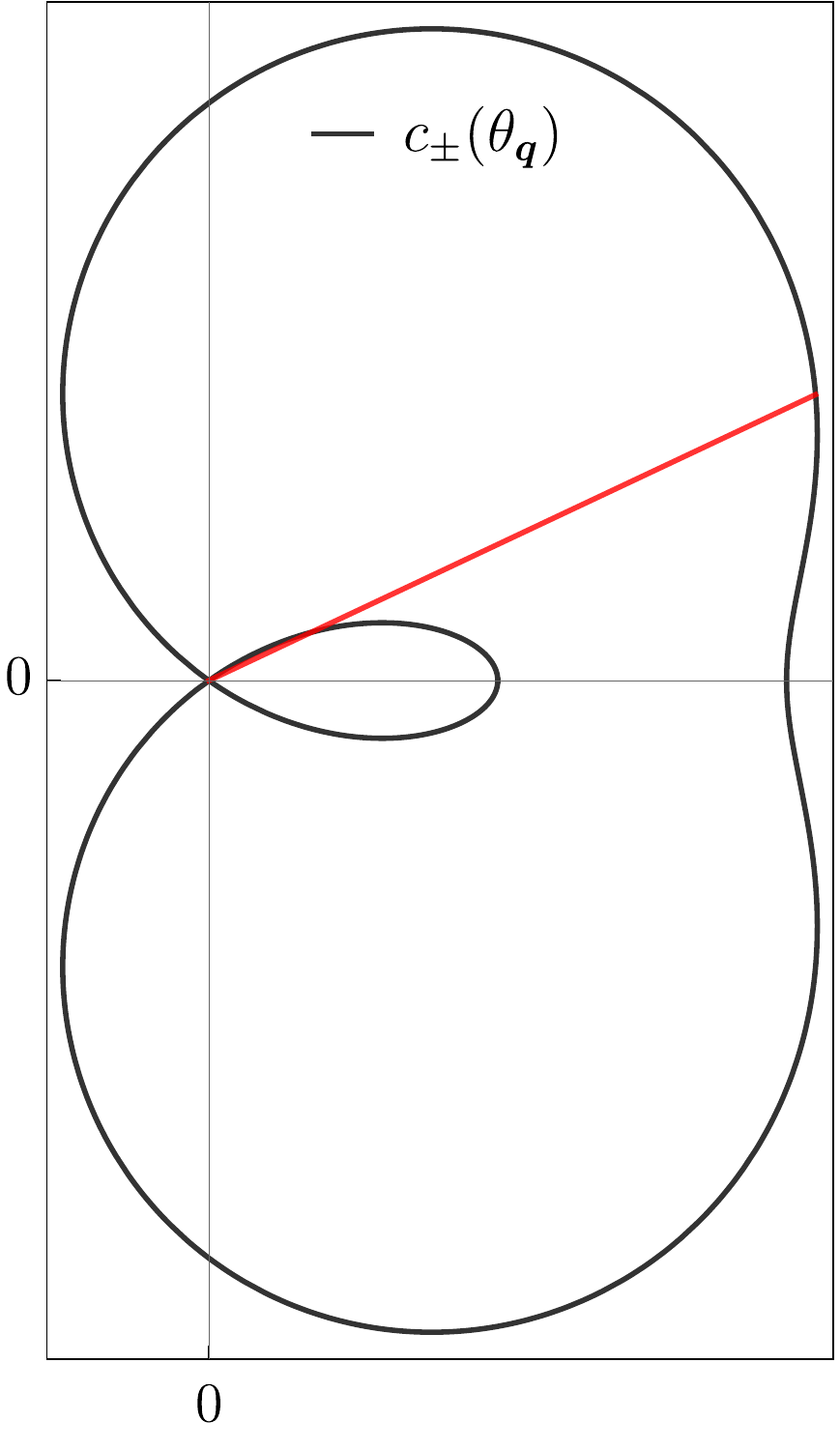}
\caption{(Color online) Polar plot of the sound speeds; the polarization points directly  to the right.  That is, the distance along a straight line line drawn from the origin and making an angle $\theta$ with the $x$-axis to its intersection with the curve is proportional to the sound speed of a mode propagating at the same angle $\theta$ to the mean polarization direction $\hx$. There are two intersections for each such line, corresponding to the two roots given in equation \eqref{cplusminus} for the sound speeds. Here we have taken $v_\rho=1$, $v_p=c_0=2$, and  
$\gamma=.3$ (all in arbitrary units).}
\label{sound speeds}
\end{figure}

We find that the polarization $\hp$,  has a simple scaling form for its spatio-temporally Fourier transformed correlation function:
\beq
C_{pp}({\bf q},\omega)\equiv\langle |p_y(\bq, \omega)|^2\rangle=\left({1\over q^2}\right)F_{pp}\bigg(\left({\omega\over q}\right), \theta_\bq\bigg)\, ,
\label{pycorscale}
\eeq
where   the scaling function $F_{pp}(u, \theta_\bq)$ is given in the ALP;  and $\theta_{\bf q}\equiv \tan ({q_y/q_x})$ is the angle between $\bq$ and  the direction $\hx$ of  the mean polarization.
The {\it positions} of the peaks in $C_{pp}(\bq,\omega)$ versus $\omega$~\cite{unreal} (but most definitely {\it not} their widths),  are precisely those found for dry active matter in \cite{tonertu95, toner98, toner05}; i.e., $\omega_{\rm peak}=c_\pm(\theta_\bq)q$, where $c_\pm(\theta_\bq)$ is   given by \eqref{cplusminus}
and  
plotted in Figure (\ref{sound speeds}).  

These peak positions agree with those found in  the experiments of \cite{Geyer17} on Quinke rollers.
 
The density-density correlation function $C_{\rho\rho}(\bq,\omega)\equiv\langle |\delta\rho(\bq, \omega)|^2\rangle$, and the density-polarization cross-correlation $C_{p\rho}(\bq,\omega)\equiv\langle p_y(\bq, \omega)
\delta\rho(-\bq, -\omega)\rangle$, both obey similar scaling laws, which are given in detail in the ALP.

Integrating these spatio-temporally Fourier-transformed correlation functions over all frequencies 
$\omega$ shows that the equal time correlation functions 
$C_{pp}(\bq)\equiv\langle |p_y(\bq, t)|^2\rangle$, $C_{\rho\rho}(\bq)\equiv\langle |\delta\rho(\bq, t)|^2\rangle$, 
and $C_{p\rho}(\bq)\equiv\langle p_y(\bq, t)\delta\rho(-\bq,t)\rangle$ all scale like $1/q$. Their dependence on the direction of $\bq$ is given explicitly in the ALP.

Fourier transforming these in space shows that the real space, equal-time correlation functions $C_{pp}(\br)=\langle p_y(\br+\bR,t)p_y(\bR,t)\rangle$, $C_{\rho\rho}(\br)\equiv\langle \delta\rho(\br+\bR,t)\delta\rho(\bR,t)\rangle$, and $C_{p\rho}(\br)\equiv\langle p_y(\br+\bR,t)\delta\rho(\bR,t)\rangle$ all scale like $1/r$, and depend on the direction of $\br$. Explicit expressions for this direction-dependence are given in the ALP.

These predictions could also be tested experimentally in systems in which the active particles can be imaged, like those of \cite{schaller13,bricard2013}. Although the anisotropy of the system ensures that all the correlators are anisotropic functions of distance $\bf r$, {\cgreen  nonetheless,} their spatial scaling remains {\em isotropic}. That is, the anisotropy exponent $\zeta$ that determines the relative scaling between $x$ and $y$ is $\zeta=1$, in contrast to the Toner-Tu model~\cite{toner98}.

The correlator $C_{\rho\rho}(\br-\br')$ can be used to obtain the result (\ref{gnf})  for the giant number fluctuations. The bulk velocity can be 
  obtained from $p_y(\br, t)$ and $\delta\rho(\br ,t)$  through the aforementioned solution of the Stokes equation subject to the active boundary condition. This  in turn allows us to derive the anomalous diffusion \eqref{anomdifshort}; see the ALP\cite{alp} for detailed derivations.

We will now provide an outline of how we obtained  
  these results. Details can be found in the ALP.




In the presence of friction from the substrate, there is no momentum 
conservation  on the surface, so  the only conserved 
variable on the surface 
is the active particle 
number. 
We also include the {\it bulk} fluid velocity $\bv(\br, t)$, which is defined throughout the semi-infinite three dimensional   (3D) space above the surface, since in that space momentum (which is equivalent to velocity in the limit of an incompressible bulk fluid) {\it is} conserved. However, we  work in the Stokesian limit, in which viscous forces dominate inertial ones.


 We  formulate the hydrodynamic equations for these variables by expanding their equations of motion phenomenologically in powers of fluctuations of both fields $\hp$ and $\rho$ from their mean values, and in spatio-temporal gradients. In doing  so, we respect all symmetries and conservation laws of the underlying dynamics.  
   In our non-equilibrium system, 
additional equilibrium constraints like detailed balance do not apply.  Our system has underlying rotational invariance in the plane of the surface, which is {\it spontaneously} broken by 
the active particles when 
they align their polarizations.

 
Conservation of the active  particles implies that  $\rho(\brp,t)$ obeys a continuity equation: 
\bea
\partial_t\rho + {\boldsymbol\nabla}_s\cdot {\bf J}_\rho &=& 0\,, \label{cont}
\eea
where ${\boldsymbol \nabla}_s\equiv {\hat{\bf x}}\partial/\partial x +
{\hat{\bf y}}\partial/\partial y$ is the 2D gradient operator, with ${\hat{\bf x}}$ and
${\hat{\bf y}}$ the unit vectors along the $x$ and $y$ axis respectively. 
We  phenomenologically expand   the active particle 
current ${\bf J}_\rho$
to leading order in powers of the bulk velocity  evaluated at the surface ${\bf v}(\brp, z=0)$, and gradients,   while respecting
rotation invariance. In practice, this means we can make the vector ${\bf J}_\rho$ only out of vectors the 
 system itself chooses, i.e., out of gradients, the surface velocity 
$\bv_s(\brp, t)\equiv{\bf v}(\brp, z=0,t)$, and the polarization $\hp(\brp, t)$. These constraints force ${\bf J}_\rho $ to take the form:
\bea
{\bf J_\rho}(\brp) &=& \rho_e(\rho, |\bv_s|) {\bf v}_s(x,y) + \kappa (\rho , |\bv_s| )  \hp
\label{currrho}
\eea
to leading order in gradients.
The factor   $\kappa (\rho, |\bv_s|)$ is an active parameter reflecting the self-propulsion of the particles through interaction with the solid substrate, 
  while the $\rho_e$ term reflects convection of the active particles by the passive fluid above them. The parameter $\rho_e\ne\rho$ in general due to drag between the active particles and the substrate.






In calculating the bulk velocity $\bv(\brp,z,t)$, we  assume  the bulk fluid is in the extreme ``Stokesian'' limit, in which inertia is negligible relative to viscous drag. 
This should be appropriate for most systems in  which the active particles are microscopic, since the Reynolds' number will be extremely low for such particles.  It is, however, certainly not valid for bottom-feeding fish, so the title of this paper takes some poetic license!

In this limit, the three-dimensional (3D) incompressible bulk velocity field ${\bf 
v}=(v_i,v_z),\,i=x,y$ satisfies 
the 3D Stokes' equation
\begin{equation}
 \eta\nabla^2_3v_\alpha (\brp,z)= \partial_\alpha \Pi (\brp,z)  ,
 \label{Stokes}
\end{equation}
 where $\eta$ is the bulk viscosity of the fluid, together with the incompressibility constraint ${\boldsymbol\nabla}_3\cdot {\bf v}=0$.   Here  ${\boldsymbol \nabla}_3\equiv {\hat{\bf x}} \partial/\partial x +
{\hat{\bf y}}\partial/\partial y+
{\hat{\bf z}}\partial/\partial z$ is the  full three-dimensional  gradient operator, with ${\hat{\bf x}}$, ${\hat{\bf y}}$,  and
${\hat{\bf z}}$ as the unit vectors along the $x$, $y$, and $z$  axes respectively, and $\Pi$ is the bulk pressure which enforces the incompressibility constraint.




This equation (\ref{Stokes}) can be solved exactly for the {\it bulk} velocity $\bv(\brp,z,t)$ in terms of the {\it surface} velocity $\bv_s(\brp,t)$. If we Fourier expand the surface velocity:
\beq
\bv_s(\brp,t)={1\over\sqrt{L_xL_y}}\sum_\bq \bv_s(\bq,t)e^{i\bq\cdot\brp}
\label{vs Fourier}
\eeq
where $(L_x,L_y)$ are the linear dimensions of our (presumed rectangular) surface, then, as we show in the ALP, 
the bulk velocity $\bv(\brp,z,t)$  is given by
\beq
\bv(\brp,z,t)={1\over\sqrt{L_xL_y}}\sum_\bq [\bv_s(\bq,t)-z(\bq\cdot\bv_s)({\hat \bq}+i{\hat \bz})]e^{-qz+i\bq\cdot\br_\perp} \,.
\label{v bulk Fourier}
\eeq

The last ingredient in our theory is the boundary condition on the bulk fluid velocity at the interface.
The active particles at the solid-liquid interface generate active forces, which change the boundary condition from  the familiar partial-slip boundary condition to:

\bew
\begin{equation}
v_{si}(\brp,t)=v_a(\rho) p_i(\brp,t) + \zeta_1(\rho)\hp\cdot\nabla_Sp_i +\zeta_2(\rho)p_i\nabla_S\cdot\hp +p_i\hp\cdot\nabla_S\zeta(\rho) + \mu\eta\bigg(\frac{\partial 
v_i(\brp,z,t)}{\partial z}\bigg)_{z=0}- \partial_iP_s(\rho)  \,,
\label{activebc}
\end{equation}
\ew
 where $v_a(\rho)$ is the spontaneous self-propulsion speed of the active particles relative to the solid substrate, $\zeta_{1,2}$ and $\zeta$ are coefficients of the active stresses permitted by symmetry, and 
$P_s(\rho)$ is a surface osmotic pressure. As before $i=(x,y)$.  For a system in thermal  equilibrium, $v_a=0=\zeta_{1,2}(\rho) = \zeta(\rho)$, and (\ref{activebc}) reduces to the well-known equilibrium partial slip boundary condition~\cite{partial-slip}.




We now turn to the equation of motion for $\hp$. As the active particles are polar, the system 
 lacks  $\hp\rightarrow
-\hp$ symmetry. This 
 allows $\partial_t\hp$ to contain terms even in $\hp$. 
 The most general equation of motion for $p_k$  allowed by symmetry, neglecting 
  ``irrelevant" terms, 
\bew
\bea
\partial_tp_k=
T_{ki}\bigg(\alpha v_{si} -\lambda_{pv}(\bv_s\cdot\nabla_s)p_i +
\left({\nu_1-1 \over 2}\right)p_j\partial_iv_{sj} 
+\left({\nu_1+1 \over 2}\right)(\hp\cdot\nabla_s)v_{si}
-\lambda(\hp\cdot\nabla_s)p_i
-\partial_iP_p(\rho)+f_{i}\bigg) ,\nonumber\\ \label{pi}
\eea
\ew
where the 
projection operator
$T_{ki}\equiv\delta^s_{ki}-p_kp_i$
insures that the fixed length condition $|\hp|=1$ on $\hp$ is preserved. 
It is the breaking of  Galilean invariance by the solid substrate that allows $\lambda_{pv}$ to differ from $1$, and the presence of the  ``self-advection" term
$\alpha$ in (\ref{pi}). 
The terms proportional to $\nu_1$ are ``flow alignment terms'', identical in form to those found in nematic liquid crystals \cite{martin1972}. The term with coefficient $\lambda$ is allowed by the polar symmetry of the particles, and can be interpreted as self advection of the particle polarity in its own direction. The function $P_p(\rho)$ is a  density dependent ``surface polarization pressure''  independent of the ``osmotic pressure'' $P_s(\rho)$ introduced earlier. 
 We have also added to the equation of motion (\ref{pi}) a white noise ${\bf f}$ with statistics
\begin{equation}
 \langle f_{ i} ({\bf r}_{_\perp},t)f_{ j} ({\bf r}_{_\perp}',t')\rangle=2 D_p\delta_{ij}\delta(
 {\bf r}_{_\perp}-{\bf r}_{_\perp}') 
\delta(t-t')\,.
\label{p noise}
\end{equation}





 Our hydrodynamic model, then, is summarized by the equations of motion (\ref{cont}), (\ref{currrho}),  and (\ref{pi}) for $\rho$  and $\hp$, respectively, and the solution (\ref{v bulk Fourier}) of the Stokes equation (\ref{Stokes}) for the bulk velocity field $\bv(x,y,z,t)$ obtained with the boundary condition (\ref{activebc}). 
 Fluctuations also involve the noise correlations  (\ref{p noise}) .


These equations of motion and boundary conditions  have an obvious spatially uniform, steady state solution:
$
\rho(\brp,t)=\rho_0 \,,
\hp(\brp,t)=\hx$,
where we have defined
$v_0\equiv v_a(\rho_0)$
and have chosen the $\hx$ axis of our coordinate system to be along the (spontaneously chosen) direction of 
polarization, as illustrated in figure (\ref{schem}).

To study fluctuations about this steady state, we expand the equations of motion (\ref{cont}), (\ref{currrho}),  and (\ref{pi}) for $\rho$  and $\hp$, and the boundary condition (\ref{activebc}), to linear order in $\delta\rho$ and $p_y$. 
We  obtain the bulk velocity $\bv(\brp,z,t)$ from the surface velocity $\bv_s(\brp,t)$ using our solution (\ref{v bulk Fourier}) of the Stokes equation.  This ultimately produces Eqs.~(\ref{rholinprl}) and (\ref{pyfin}) , where the phenomenological hydrodynamic parameters $v_\rho$, $v_p$, $\gamma$, $\gamma$, $\rho_c$, and $\sigma_t$ are all related to  the expansion coefficients of the various parameters introduced above when expanded in powers of the small fluctuations $\delta\rho$ and   $p_y$.
The rather involved details of this calculation are given in the ALP.  

 The correlation functions can be straightforwardly determined from these equations of motion, and shown to have peaks at $\omega_{\rm peak}=c_\pm(\theta_\bq)q$, where $c_\pm(\theta_\bq)$ is  given by 
\begin{eqnarray}
c_{\pm}\left(\theta_\bq \right) &=&
\pm \sqrt{{1 \over 4}\left(v_\rho -v_p\right)^2 \cos^2
\theta_\bq + c^2_0
\sin^2 \theta_\bq} \nonumber \\ 
&& +\left({v_\rho + v_p \over 2}\right)\cos\theta_\bq \quad \,.
\label{cplusminus}
\end{eqnarray}

We have presented a comprehensive hydrodynamic theory of flocking at a solid-liquid interface. This theory makes quantitative , experimentally testable predictions about orientational long range order, spatio-temporal scaling of fluctuations, giant number fluctuations and anomalous diffusion along directions parallel to the solid-liquid interface. These predictions are {\em exact} in the asymptotic long wavelength 
limit,  as {\cgreen will be shown in the ALP } using renormalization group arguments.
 One simple 
  variant on our system  would be to replace the bulk isotropic fluid of our system with  a nematic.

Acknowledgements: One of us (AB) thanks 
the SERB, DST (India) for partial financial support through the MATRICS scheme [file no.: MTR/2020/000406]. NS is partially supported by Netherlands Organization for Scientific Research (NWO), through the Vidi grant No. 2016/N/00075794. We thank S. Ramaswamy for sharing reference \cite{maitra2018} with us.  NS thanks Institut Curie and MPIPKS for their   support through postdoctoral fellowships while some of this work was being done.  AB thanks the MPIPKS, Dresden for their hospitality, and their support through their Visitors' Program, while a portion of this work was underway. JT likewise thanks the MPIPKS for their hospitality, and their support through the Martin Gutzwiller Fellowship, and the Higgs Center of the University of Edinburgh for their support with a Higgs Fellowship.


\begin{thebibliography}{7}
\bibitem{xtalfoot} Exceptions to this result are two dimensional crystals (see, e.g., \cite{2dxtal})
and fluctuating tethered membranes (see, e.g., \cite{teth}).
\end{thebibliography}
\end{document}